# All-Electrical Self-Switching of van der Waals Chiral Antiferromagnet

Junlin Xiong,[1,*] Jiawei Jiang,[1,7,*] Yanwei Cui,[1,*,†] Han Gao,[3,*], Ji Zhou,[1] Zijia Liu,[4] KuiKui Zhang,[5] Shaobo Cheng,[5] Kehui Wu,[4] Sang-Wook Cheong,[6] Kai Chang,[7] Zhongkai Liu,[4,‡] Hongxin Yang,[7,§] Shi-Jun Liang,[1] Bin Cheng[1,2,¶] and Feng Miao[1,8,‡‡]

[1]Institute of Brain-Inspired Intelligence, National Laboratory of Solid State Microstructures, School of Physics, Collaborative Innovation Center of Advanced Microstructures, Nanjing University, Nanjing 210093, China.

[2]Institute of Interdisciplinary Physical Sciences, School of Science, Nanjing University of Science and Technology, Nanjing 210094, China.

[3]School of Physical Science and Technology, ShanghaiTech Laboratory for Topological Physics, ShanghaiTech University, Shanghai 201210, China.

[4]Institute of Physics, Chinese Academy of Sciences, Beijing 100190, China.

[5]Henan Key Laboratory of Diamond Optoelectronic Materials and Devices, Key Laboratory of Material Physics, Ministry of Education, School of Physics, Zhengzhou University, Zhengzhou 450052, China.

[6]Center for Quantum Materials Synthesis and Department of Physics and Astronomy, Rutgers, The State University of New Jersey, Piscataway, NJ, 08854, USA.

[7]Center for Quantum Matter, School of Physics, Zhejiang University, Hangzhou 310058, China.

[8] Jiangsu Physical Science Research Center, Nanjing 210093, China.

*Abstract*—Antiferromagnets have garnered significant attention due to their negligible stray field and ultrafast magnetic dynamics, which are promising for high-density and ultrafast spintronic applications. Their dual functionality as both spin sources and information carriers could enable all-electrical self-induced switching of antiferromagnetic order, offering great potential for ultra-compact spintronic devices. However, related progress is still elusive. Here, we report the deterministic switching of chiral antiferromagnetic orders induced by charge current at zero external magnetic field in the van der Waals (vdW) magnetically intercalated transition metal dichalcogenide $CoTa_3S_6$. This system exhibits strong interactions between cobalt atom magnetic moment lattice and itinerant electrons within the metallic layers, as demonstrated by temperature-dependent angle-resolved photoemission, scanning tunneling spectroscopy, and topological Nernst effect measurements. Notably, the itinerant-localization interactions lead to current-induced chiral spin orbit torques as well as Ruderman-Kittel-Kasuya-Yosida (RKKY) exchange torques that interact with the localized magnetic moments, facilitating all-electrical switching of the chiral magnetic order in the $CoTa_3S_6$ flake. Our work opens a promising avenue for manipulating antiferromagnetic orders by delicately engineering the synergistic interactions between magnetic moments and itinerant electrons.

___________________________________________

*These authors contributed equally to this work.
†Contact author: ywcui@nju.edu.cn
‡Contact author: liuzhk@shanghaitech.edu.cn
§Contact author: hongxin.yang@zju.edu.cn
¶Contact author: bincheng@njust.edu.cn
‡‡Contact author: miao@nju.edu.cn

Exploring novel quantum antiferromagnetic (AFM) materials and realizing electrical manipulation of the AFM orders represent a key frontier in condensed-matter physics and materials science [1-10], with promising implications for the next generation of ultrafast, miniaturized spintronic devices [11-15]. Of particular interest are topological AFM systems that host non-collinear spin textures [16], which can offer an alternative to ferromagnets for digital information storage due to their negligible stray fields and ultrafast dynamics. A central feature of non-collinear antiferromagnets (NCAFs) is their spin chirality, which gives rise to a plethora of exotic intrinsic properties, including anomalous or topological Hall and Nernst effects [17-22]. This enables efficient electrical readout of magnetic states despite the absence of any net magnetization. Together with their ability to switch Nèel vectors using current-induced spin-orbit and spin-transfer torques from heavy metals, NCAFs serve as spin information carriers in spintronic devices [23-30]. In addition, NCAFs can also act as spin sources, exhibiting unconventional charge-to-spin conversion effects [31-36], where the spin polarization occurs in directions inaccessible through conventional spin Hall effects. Such dual functionality, as both spin source and information carrier, has the potential for realizing self-induced electrical switching of antiferromagnetic order, a mechanism that could enable highly efficient and compact quantum spintronic devices, eliminating the need for complex multilayered heterostructures. Despite these potentials, effective strategies for achieving self-induced electrical switching within a single NCAF material remain elusive, hindering the exploitation of these NCAF materials for the development of memory and computing quantum devices.

In this work, we demonstrate the current induced all-electrical switching of chiral magnetic order in van der Waals (vdW) magnetically intercalated transition metal dichalcogenide $CoTa_3S_6$. The crystal structure of $CoTa_3S_6$ consists of a chiral, alternating stacking of cobalt atom lattice carrying localized magnetic moments, and metallic 2H-TaS$_2$ layer hosting itinerant electrons [Fig. 1(a)]. The interactions between these magnetic moments and the itinerant electrons at the low temperature, which lead to the emergence of topological hybridization bands, was confirmed by temperature dependent scanning tunneling spectroscopy (STS), angle-resolved photoemission spectroscopy (ARPES) and topological Nernst effect (TNE) measurements. Remarkably, we demonstrate that the interplay of interactions between itinerant electrons and magnetic moments facilitates the deterministic and reversible all-electrical switching of chiral magnetic order through the application of a current pulse. In this process, the spin-orbit torque interaction switches each cobalt magnetic moment individually, followed by a collective rotation driven by the Ruderman–Kittel–Kasuya–Yosida (RKKY) magnetic exchange interaction. Our work opens a new avenue for switching the topological antiferromagnetic orders by precisely tailoring the interactions between localized magnetic moments and itinerant electrons.

$CoTa_3S_6$ single crystals were synthesized by the chemical vapour transport (see Supplementary Note 1 and Fig. S1 [37] for details on the synthesis process and chemical composition analysis). The crystal structure of $CoTa_3S_6$ was characterized using scanning transmission electron microscopy (STEM), which shows the atomic-scale flatness of each constituent layer [Fig. 1(b)]. The intercalated magnetic cobalt atom lattice breaks the mirror and inversion symmetry of 2H-TaS$_2$, forming a non-coplanar antiferromagnetic order [21,22]. This chiral magnetic order [see Fig. 1(c) and (d)] can create a fictitious perpendicular magnetic field on itinerant electrons [49-51], thus generating a large spontaneous topological Hall effect (THE).

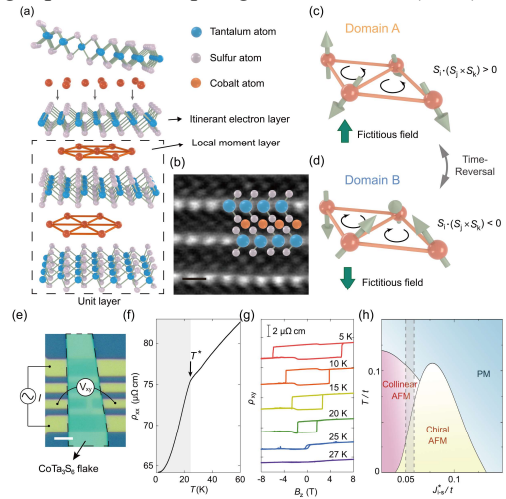

FIG. 1. (a) Schematic of crystal structure of $CoTa_3S_6$. (b) Cross-sectional STEM image of a typical $CoTa_3S_6$ flake. The scale bar is 0.5 nm. (c)-(d) Schematic of the atomic-scale chiral magnetic order with positive (c) or negative (d) scalar chirality. (e) Optical image of a typical device. The scale bar is 4 μm. (f) Temperature dependence of the longitudinal resistivity. The grey region represents the chiral magnetic phase. (g) Hall resistivity as a function of the perpendicular magnetic field. (h) Magnetic phase diagram calculated from tight-binding model, indicating the emergence of chiral AFM state.

To experimentally characterize the magnetic properties of $CoTa_3S_6$, we fabricated devices with several pairs of Hall bars using exfoliated $CoTa_3S_6$ samples [see the typical optical image in the Fig. 1(e)] for transport measurements. We first carried out measurements of the longitudinal ($\rho_{xx}$) and Hall ($\rho_{xy}$) resistivities under different temperatures. When the temperature decreases, $\rho_{xx}(T)$ decreases in a nearly linear trend before a Fermi liquid behavior with a $T^2$ dependence below $T^* \approx 25$ K [Fig. 1(f)], where CoTa$_3$S$_6$ undergoes a magnetic phase transition to chiral magnetic state. Meanwhile, at low temperatures, we observe a square-shaped magnetic hysteresis loop in $\rho_{xy}$ [Fig. 1(g)], confirming the spontaneous topological hall effect induced by the chiral magnetic order in CoTa$_3$S$_6$. This hysteresis loop gradually vanishes as the temperature increases, indicating a transition temperature of approximately 25 K, consistent with previous reports [21,22].

We then investigate the underlying mechanism of the chiral magnetic phase in CoTa$_3$S$_6$ using density functional theory (DFT) and a tight-binding (TB) model, as well as dynamical mean-field theory (DMFT) to capture finite-temperature effects (see Supplementary Note 2 [37] for more details). For a small $J_{im}/t$, the magnetic phase diagram [Fig. 1(h)] derived from the TB model predicts a collinear antiferromagnetic (AFM) phase. Here, $J_{im}$ and $t$ represent the coupling between itinerant electrons and magnetic moments, and the electron hopping, respectively. As $J_{im}/t$ increases, a non-trivial chiral magnetic phase emerges at low temperatures. By comparing the TB band structures with DFT-calculated results, we extract a coupling value of $J_{im}^*/t = 0.050 - 0.057$, indicating that the ground state of CoTa$_3$S$_6$ lies within the chiral magnetic phase of the phase diagram, consistent with the experimental observations. At higher temperatures, the chiral magnetic phase transitions into a collinear AFM phase and eventually into a paramagnetic (PM) phase. This temperature-dependent behavior, calculated using the TB model, aligns with previous experimental observations [21,22], where a collinear antiferromagnetic order appears below the Néel temperature, and a magnetic transition to chiral magnetic phase is observed at $T^* \approx 25$ K, indicating the crucial role of interaction $J_{im}$ in the formation of chiral magnetic order in CoTa$_3$S$_6$.

We next performed angle-resolved photoemission spectroscopy (ARPES) and scanning tunneling spectroscopy (STS) measurements to substantiate the hybridization between itinerant electrons and magnetic moments in the CoTa$_3$S$_6$. Figure 2(a) shows ARPES band dispersions at 12.8 K along $\bar{\Gamma} - \bar{K} - \bar{M}$ in the hexagonal surface Brillouin zone (see Fig. S3 [37]). A band located near the Fermi level ($E_F$) and close to the $\bar{K}$ point is clearly observed, consistent with the calculated Co–Ta orbital hybridization in this momentum region (see Fig. S12 [37]). The suspected hybridization band is further characterized by the density of states (DOS) peak in the energy distribution curve at $\bar{K}$ point associated with this band [marked by the red line in Fig. 2(a)], as shown in Fig. 2(b). We then measured the temperature dependence of the DOS peak intensity to investigate the thermodynamic evolution of band. As the temperature decreases from 40 K, the integrated intensity of the DOS peak remains nearly constant, followed by a modest increase below 30 K. This intensity eventually saturates around 25 K [Fig. 2(c)], which is consistent with the magnetic transition temperature $T^*$ and suggests the temperature at which the suspected hybridization state is established.

It's worth noting that, due to the experimental resolution, direct spectroscopic observation of the predicted ~10 meV gap and reconstruction of the electronic structure cannot be clearly observed. To further confirm the band hybridization and hybridization temperature of CoTa$_3$S$_6$, we carried out the STS measurements. The corresponding results at different temperatures are shown in Fig. 2(d). We observe two nearly symmetric peaks in the scanning tunneling spectrum with respect to the Fermi level appearing at 17 K, indicating the presence of the hybridization gap in CoTa$_3$S$_6$. With increasing temperature, these peaks gradually diminish and eventually vanish around 25 K, again consistent with the transition temperature $T^*$ observed in transport measurements. These results indicate that the interaction between local moments and itinerant electrons gives rise to a strongly hybridized electronic structure in the chiral magnetic phase, featuring a hybridization gap opening near the $\bar{K}$ point, as predicted by our theoretical calculations.

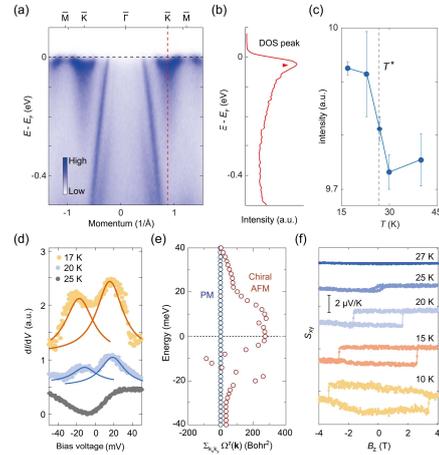

FIG. 2. (a) Angle-resolved photoemission spectroscopy spectra of CoTa$_3$S$_6$ along $\bar{\Gamma} - \bar{K} - \bar{M}$ path measured at $T = 12.8$ K. The black line represents the Fermi level. (b) Energy distribution curve along the red vertical line shown in (a). The red triangle marks the density of state (DOS) peak near the Fermi level. (c) Temperature dependence of the intensity of the DOS peak, indicating an electronic transition near the magnetic ordering. (d) Temperature evolution of the d$I$/d$V$ spectra from 17 K to 25 K. The spectra are vertically offset for clarity. (e) Sum of Berry curvature in the momentum space as a function of the energy, illustrating the topological distinction between the PM and chiral AFM phases. (f) Magnetic field dependence of Nernst signals.

Remarkably, the reduced electronic bandwidth and the resulting enhancement of the DOS in the hybridization band, driven by the chiral magnetic order, could give rise to a large Berry curvature near $E_F$, as supported by the calculated results shown in Fig. 2(e) (see Supplementary Note 1 [37] for details). To experimentally confirm the presence of this topological hybridization band near $E_F$, we measured the spontaneous topological Nernst effect (TNE) in $CoTa_3S_6$ by fabricating a thermoelectric device [see Fig. 2(f) for results at different temperatures]. Distinct from charge transport, thermoelectric signals involve the energy derivative of the Fermi-Dirac distribution and Berry curvature, and thus can serve as sensitive probes of the topological properties of Bloch electrons at Fermi surface [52,53]. At low temperatures, we observe a square-shaped hysteretic loop, indicating the spontaneous TNE [Fig. 2(f)] and supporting the existence of a topological band near $E_F$. Moreover, the presence of a finite Nernst signal at zero external magnetic field and vanishing magnetization (see Fig. S7 [37]) strongly indicates a Berry-curvature-driven origin, rather than a trivial magnetization-induced effect. As the temperature rises, the hysteresis loop fades and eventually vanishes around 25 K, again aligning with the transition temperature $T^*$. These observations suggest that the chiral magnetic order modifies the band structure via hybridization between itinerant electrons and localized magnetic moments, in agreement with the calculated band evolution shown in Fig. S2 [37].

The interactions between itinerant electrons in the topological bands and localized magnetic moments in $CoTa_3S_6$ allow for the electrical control of magnetic moments under an applied current, facilitating a write operation of the antiferromagnetic state. We first demonstrated the current-induced nonvolatile and deterministic switching of the chiral magnetic order in the $CoTa_3S_6$ samples under various external magnetic fields (denoted as $B_\parallel$) aligned with the direction of charge current. We set $B_\parallel$ = -0.1 T, and applied a pulsed current, sweeping the corresponding current density $J_{pulse}$ from $-10$ to 10 MA/cm$^2$, while monitoring the Hall resistance [Fig. 3(a)]. Specifically, sweeping $J_{pulse}$ up or down to a critical value of approximately $6.6 \times 10^6$ A cm$^{-2}$ reverses the sign of the Hall resistance, which reflects the scalar chirality of the magnetic "skyrmion" order. The polarity of this current-induced switching was observed to be anticlockwise in this case. Notably, as $B_\parallel$ was set to zero [Fig. 3(b)], we still observed the same switching polarity as above, realizing field-free electrical switching of the chiral magnetic order by increasing the current density to about $7 \times 10^6$ A cm$^{-2}$. This critical current density required to switch such non-collinear antiferromagnetic (NCAF) order in single-crystal flakes of $CoTa_3S_6$ is comparable with that state-of-the-art value for switching magnetic order in multilayered magnetic heterostructures [23,24,27,54-56]. Additionally, the anticlockwise switching polarity is retained at $B_\parallel$ = 0.1 T [Fig. 3(c)] and eventually reversed at $B_\parallel$ = -0.2 T [Fig. 3(d)]. Notably, the magnetic field dependent switching polarity is also observed in chiral antiferromagnet $Mn_3Sn$/heavy metal heterostructures [24,26-28], but unlike our system, the switching polarity in their systems reversed at zero magnetic field. Our results show that $CoTa_3S_6$ single crystals can uniquely serve as both a spin-torque source and a spin-information carrier, thereby facilitating the development of ultra-compact spintronic devices.

We then investigated the deterministic and reversible switching of the chiral magnetic order between two states with opposite scalar chirality over multiple cycles [see Fig. 3(e) and (f)]. Successive 1-ms writing pulses of amplitude $8 \times 10^6$ A cm$^{-2}$ were applied alternately to the $CoTa_3S_6$ sample. The first current pulse reliably reversed the Hall resistance, while the subsequent two pulses had no observable effect, indicating a stable switching process. Additionally, by reducing the current pulse width from 1 ms to 0.2 ms, we were still able to deterministically reverse the chiral magnetic order with pulses of the same amplitude. These observations indicate that the switching is dominated by spin-torque-related mechanisms, although thermal effects may assist the process.

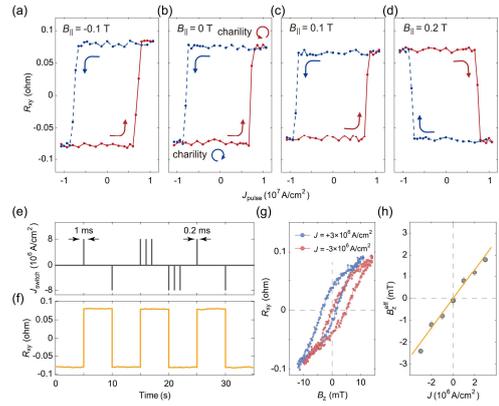

FIG. 3. (a)-(d) Current-induced switching of chiral magnetic order under $B_\parallel$ = -0.1 T (a), $B_\parallel$ = 0 T (b), $B_\parallel$ = 0.1 T (c) and $B_\parallel$ = 0.2 T (d). The duration of the current pulses is 200 μs, and all experiments were conducted at a temperature of 20 K. (e) Deterministic switching by a series of current pulses applied in the device. (f) The resulting Hall resistance is measured using an a.c. excitation current of 5 μA. (g) The measured THE hysteresis loops under the perpendicular magnetic fields while a positive or negative direct charge current is applied, revealing a polarity-dependent loop shift. (h) The effective field $B_z^{eff}$ measured as a function of charge current, confirming a spin-torque origin.

We then further examine the presence of current-induced effective field $H_{\text{eff}}^z$ by performing the THE loop shift measurements [54,57-59]. The Hall resistance was measured as a function of the perpendicular magnetic field for applied positive and negative direct currents, with the corresponding results shown in Fig. 3(g). The center of the hysteresis loop shifts to the left or right, depending on the direction of current. The experimental results repeated for several different current magnitudes, summarized in Fig. 3(h), show that the effective perpendicular field $H_{\text{eff}}^z$, extracted from loop shifts, exhibits a nearly linear dependence on the applied direct current $J$. These observations confirm the presence of a current-induced out-of-plane effective field, which could facilitate the all-electrical switching of the NCAF order, which will be discussed later.

Finally, we offer insights into the observed current-induced switching of the atomic-scale chiral magnetic order. The broken mirror and inversion symmetry in the CoTa$_3$S$_6$ could facilitate an unconventional charge-spin conversion effect, generating a spin current carried by itinerant electrons that acts on the magnetic moments. First-principles calculations reveal that this effect originates from the strong spin Berry curvature (SBC) enhancement near the Fermi level, particularly within a narrow hybridization band formed by the coupling between localized Co 3d and itinerant Ta 5d electrons (see Supplementary Note 3 [37] for details). As shown in Fig. 4 (a) and (b), substantial SBC is present on the topological hybridization band, leading to a non-zero spin conductivity $\sigma_{xx}^z$ near the Fermi level. This large $\sigma_{xx}^z$ indicates that a charge current applied to CoTa$_3$S$_6$ flake would induce an imbalance in the distribution of electrons carrying spin polarization, leading to the generation of spin accumulation with an out-of-plane component $s_z$. This current induced z-spin polarization could generate an effective field acting on the chiral antiferromagnetic moments [21,22,60] in the CoTa$_3$S$_6$, consistent with the experimental observations in Fig. 3(g). To support this interpretation, we performed atomistic LLGS simulations that model the magnetization dynamics under current-induced spin accumulation, which reproduce the observed field shift (~0.8 mT per $10^6$ A/cm$^2$), validating the torque-based switching mechanism (see Fig. S9 [37]). Furthermore, the current-induced z-spin accumulation further interacts with magnetic moments of cobalt atoms, generating chiral spin-orbit torques that enable the switching of the scalar chirality of the NCAF order without the need for an external magnetic field (see Supplementary Note 4 [37] for details). This process of current-induced switching is depicted in Fig. 4(c)-(e), where the spin dynamics simulation, based on the atomistic Landau-Lifshitz-Gilbert (LLG) equation, demonstrates two stages of chirality evolution [Fig. 4(d)].

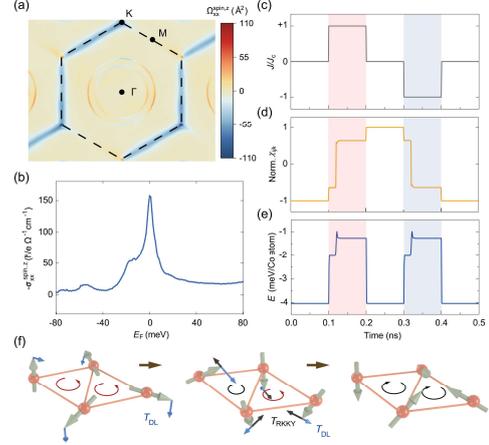

FIG. 4. (a) Calculated k-resolved spin Berry curvature in CoTa$_3$S$_6$. The color code indicates the magnitude of the spin Berry curvature. (b) Calculated spin conductivity $\sigma_{xx}^{\text{spin},z}$ as a function of the Fermi energy, showing strong enhancement near the hybridization gap. (c) The input charge current pulse in the atomistic spin dynamics simulations. (d) The time evolution of the scalar spin chirality $\chi_{ijk} = \boldsymbol{S}_i \cdot (\boldsymbol{S}_j \times \boldsymbol{S}_k)$ for each magnetic moment $\boldsymbol{m}_i$. Note that $J/J_c$ and normalized $\chi_{ijk}$ in (c)-(d) are dimensionless quantities. (e) The time evolution of the total energy per Co atom. (f) Schematics of the switching mechanisms in the atomic-scale chiral magnetic order.

Initially, the chiral spin-orbit torque $T_{\text{DL}} \sim \boldsymbol{m}_i \times (\boldsymbol{s}_z \times \boldsymbol{m}_i)$ switches each cobalt magnetic moment $\boldsymbol{m}_i$ ($i$ = 1, 2, 3, 4) individually. This process is followed by a collective rotation, driven by the RKKY-mediated interaction between the magnetic moments. These stages of chirality evolution during the switching process, accompanied by a sharp increase in energy, are depicted in the theoretical calculations [Fig. 4(e)]. Upon removal of the charge current, the system relaxes into a stable NCAF order with reversed chirality, as indicated by a rapid decrease in energy. Besides, when a opposite current is applied, the spin dynamics simulation demonstrates that the chirality of the NCAF order reverses again, consistent with our experimental results. Furthermore, LLGS simulations show that the critical current density is governed by the spin-orbit torque efficiency, with the simulated value (~8×10$^6$ A/cm$^2$) matching well with the experimental result (~7×10$^6$ A/cm$^2$) (see Fig. S10 [37]). We finally conceptually illustrate the interplay of the RKKY exchange torques and chiral spin-orbit torques in Fig. 4(f), which enable the self-induced switching of antiferromagnetic order without the assistance of an additional spin source layer. While our demonstrated proof-of-concept switching operated at low temperature in CoTa$_3$S$_6$, the same mechanism could apply to other chiral antiferromagnets with higher ordering temperatures, offering a potential pathway toward spintronic applications.

*Acknowledgments*—This work was supported in part by the National Natural Science Foundation of China (12322407, 62122036, 62034004, 61921005, 62271450), the National Key R&D Program of China under Grant 2023YFF0718400, 2023YFF1203600 and 2022YFA1405102, the Leading-edge Technology Program of Jiangsu Natural Science Foundation (BK20232004), the Natural Science Foundation of Jiangsu Province (No. BK20233001), the Strategic Priority Research Program of the Chinese Academy of Sciences (XDB44000000), Innovation Program for Quantum Science and Technology. F.M. and S.-J.L. would like to acknowledge support from the AIQ Foundation and the e-Science Center of Collaborative Innovation Center of Advanced Microstructures. Z. K. Liu acknowledges the support from the National Nature Science Foundation of China (92365204, 12274298) and the National Key R&D program of China (Grant No. 2022YFA1604400/03). The microfabrication center of the National Laboratory of Solid State Microstructures (NLSSM) is acknowledged for their technique support. Crystal growth at Rutgers was supported by the center for Quantum Materials Synthesis (cQMS), funded by the Gordon and Betty Moore Foundation's EPiQS initiative through grant GBMF6402, and by Rutgers University. We thank Pingfan Gu and Yu Ye for fruitful discussions in magneto-optical circular dichroism measurements.